# Mapping Responsibility Attribution in the Grenfell Tower Inquiry: A Network Analysis

Mariano Devoto[1,3] & Pablo A. Cipriotti[2,3]

1- Universidad de Buenos Aires, Facultad de Agronomía, Cátedra de Botánica General.

2- Universidad de Buenos Aires, Facultad de Agronomía, IFEVA – Departamento de Métodos Cuantitativos y Sistemas de Información

3- Consejo Nacional de Investigaciones Científicas y Técnicas (CONICET), Argentina

## Abstract

After major disasters, formal inquiries become arenas where responsibility is publicly contested. While extensive research has examined blame attribution through qualitative and actor-centred approaches, the relational structure of blame within formal accountability processes remains poorly understood. Using evidence from the Grenfell Tower Inquiry, this study analyses the web of blame presented during the Phase 2 closing proceedings, in which Counsel to the Inquiry synthesised how core participants publicly attributed responsibility to one another.

We represent this synthesis as a directed network and examine its structural properties using standard tools from network analysis. The resulting configuration is interconnected, with pronounced reciprocity and local clustering, indicating that responsibility claims were articulated within a dense institutional environment rather than as isolated, one-to-one accusations. Comparisons with neutral benchmark models show that several observed features depart from expectations based on simple structural constraints alone, revealing patterned organisation in the public articulation of blame within the Inquiry.

By applying network-analytic methods to an institutionalised representation of blame attribution, this study provides a systematic structural account of responsibility relations in a major public inquiry. The findings contribute to research on crisis governance and blame avoidance by demonstrating how accountability processes generate patterned distributions of responsibility that reflect institutional arrangements and governance contexts.

## Introduction

Large-scale disasters often generate intense struggles over responsibility. When catastrophic events unfold under conditions of uncertainty, complexity, and institutional fragmentation, questions of who is to blame rarely admit straightforward answers.

Instead, crises tend to produce competing narratives of causation, responsibility, and failure, articulated by a diverse set of actors seeking to defend their legitimacy, protect their authority, or shape subsequent accountability processes. In contemporary societies characterised by complex technological systems and manufactured risks, such struggles over blame are not exceptional but constitute a core feature of governance in crisis situations (Beck, 1992).

Research on crisis governance highlights that crises confront public authorities with profound uncertainty, time pressure, and high stakes, while simultaneously exposing institutional arrangements to intense public scrutiny (Ansell & Boin, 2019; Boin & Lodge, 2021). Under these conditions, governance extends beyond technical problem-solving to encompass the management of responsibility, legitimacy, and public narratives. In some cases, crises escalate into what have been described as institutional crises, in which the integrity and legitimacy of an entire governance system are called into question rather than the performance of a single organisation or policy (Ansell et al., 2016).

Within this literature, blame has been conceptualised as a relational and strategic phenomenon. Rather than being passively assigned, blame is actively managed through what Hood (2011) terms blame games, in which actors seek to avoid being blamed while redirecting responsibility toward others through presentational, organisational, and policy strategies. These dynamics unfold across organisational boundaries and involve multiple actors simultaneously blaming, deflecting, or reframing responsibility. Importantly, such processes do not require coordination or shared intent: patterned outcomes may emerge from decentralised, self-protective behaviour operating under common institutional pressures.

Blame attribution is also shaped by asymmetries of power and visibility. Power operates not only through overt decision-making but also through the capacity to shape agendas, define what is contestable, and render certain actors or issues invisible (Lukes, 2005). In the context of crisis inquiries, some organisations and individuals may be structurally positioned to attract scrutiny and blame, while others remain peripheral or shielded from attention. Understanding blame attribution therefore requires attention not only to individual statements, but also to the broader structure of relations through which responsibility is publicly articulated.

The 2017 Grenfell Tower fire in London constitutes a paradigmatic case of an institutional crisis. Beyond the immediate loss of life, the disaster exposed long-standing failures in housing regulation, building safety, local governance, and public oversight. In response, the Grenfell Tower Inquiry was established as a formal arena for investigating causes, responsibilities, and failures. As part of its proceedings, the Inquiry explicitly assembled and presented a *web of blame*, mapping how different core participants attributed responsibility to one another in their closing submissions. This web was not intended to establish legal liability, but to document and compare patterns of blame attribution within a structured and institutionalised setting.

In this paper, we analyse the structure of this web of blame using a network perspective. We treat the network strictly as a synthesis of public claims made within a specific institutional setting: a structured dataset of who publicly blamed whom during the Inquiry's closing phase. Our analysis is therefore descriptive and structural. We do not use network metrics to infer hidden strategies, private intentions, or the causal efficacy of blame-shifting. Instead, we ask two specific questions: First, what are the measurable structural properties of this institutionalised representation of blame (e.g., how concentrated, reciprocal, or clustered are the accusations)? Second, to what extent can these observed properties be explained by simple compositional constraints, as opposed to more patterned organization? Network analysis provides tools to characterize these structural features (Wasserman & Faust, 1994; Newman, 2010) and to compare the observed structure with neutral benchmark models. Such comparisons help identify which features depart from random expectations, but they do not test causal mechanisms or reveal strategic intent.

This study addresses two related objectives. First, it aims to characterise the structure of blame attribution within the web of blame presented during the Grenfell Tower Inquiry, examining how responsibility claims are distributed across actors, whether blame tends to be concentrated or dispersed, and how blame relationships are patterned in terms of directionality and local organisation. Second, the study assesses whether the observed structural features of this blame attribution network differ from what would be expected under simple structural constraints, using neutral benchmark models as points of comparison.

By combining insights from crisis governance and blame-avoidance theory with network analysis, this study contributes to a growing literature on accountability and governance under conditions of uncertainty. Empirically, it provides a systematic structural analysis of blame attribution in a major public inquiry. Conceptually, it demonstrates how network approaches can be used to study blame as a relational governance phenomenon while maintaining clear epistemic limits on inference. More broadly, the paper illustrates how institutional arenas of crisis inquiry generate patterned distributions of responsibility that shape post-crisis accountability debates.

## Methods

### *Data and institutional context*

The empirical material analysed in this study derives from the Grenfell Tower Inquiry, a statutory public inquiry established under the Inquiries Act 2005 to investigate the causes and circumstances of the 2017 Grenfell Tower fire. The analysis focuses on the network of blame attribution presented by Counsel to the Inquiry during the overarching closing session of Phase 2 on 10 November 2022. During this session, Counsel presented a structured synthesis of blame attributions derived from the closing submissions previously delivered by core participants over the course of the Phase 2 proceedings, as documented in the official Inquiry transcript and visualised in the 'web of blame' diagram (Grenfell Tower Inquiry, 2024). The selection and presentation of these relationships by Counsel was guided by the Inquiry's procedural aim of neutrally mapping "who blames whom and for what" to aid the assessment of causation and culpability, rather than to determine legal liability.

The analysis therefore draws exclusively on materials from the second phase of the Inquiry, which focused on responsibility, causation, and institutional failure, rather than on the first phase, which examined the events of the night of the fire. The data represent a single institutional moment corresponding to the culmination of the Inquiry's evidentiary phase. The resulting network thus constitutes a static snapshot of publicly articulated blame relations at that point in time, providing the empirical basis for characterising the structure of blame attribution within the Inquiry. The network excludes certain core participants, such as specialist subcontractors and the London Fire Brigade, not due to their irrelevance, but because their formal positions, as synthesised by Counsel, did not contribute centrally to the reciprocal, multi-party blame structure that the session sought to illustrate. This boundary reflects the Inquiry's own synthesis for illustrative purposes; a different boundary (including, e.g., victim groups or other peripheral actors) would produce a different network structure for analysis.

### *Network construction and definition of blame attribution*

For the purposes of this study, a blame attribution is defined as an explicit statement made in a formal closing submission in which one actor identifies another actor as bearing responsibility, causative relevance as claimed by the actor, or culpability for aspects of the Grenfell disaster. Blame attributions include both direct assertions of fault and indirect forms of responsibility assignment, such as claims that another actor's decisions, omissions, or practices materially contributed to the outcome.

This operationalisation follows the Inquiry's own approach, which treated blame attribution as a descriptive mapping of responsibility claims articulated within a formal accountability process. The analysis does not assess the validity, accuracy, or normative justification of any blame attribution, nor does it attempt to infer private beliefs, intentions, or strategic motivations of the actors involved. The scope of included attributions is delimited by the synthesis presented by Counsel, which focused on the most prominent and materially relevant claims among the main core participants.

The blame attribution network is represented as a directed graph (Wasserman & Faust, 1994; Newman, 2010). Nodes correspond to organisational or institutional actors identified as core participants in the Inquiry, including public authorities, private companies, professional firms, and regulatory organisations. Individual persons are not treated as separate nodes unless they were explicitly framed as independent responsibility-bearing actors within the Inquiry's synthesis. Node inclusion is therefore determined by an actor's formal role in the proceedings and their participation in the documented structure of mutual blame.

Directed edges represent blame attributions. A directed edge from actor *i* to actor *j* indicates that *i* publicly attributed responsibility to *j* in its closing submission, as captured and synthesised by Counsel to the Inquiry. Multiple blame attributions between the same pair of actors are represented as a single directed edge, reflecting the presence rather than the frequency or intensity of blame. Self-attributions of responsibility are excluded from the network. This representation provides the formal basis for examining how blame relationships are distributed and patterned across actors.

To aid in visual interpretation and descriptive analysis, each node was assigned to a *guild* based on its primary functional role within the Grenfell Tower refurbishment and regulatory system. The four guilds, reflected in the node colors of Figure 1, are: Product Manufacturer (e.g., Arconic, Celotex), Government Body (e.g., Government/DCLG, RBKC), Construction (e.g., Harley Facades, Rydon), and Fire Safety/Engineering (e.g., Exova). This categorization is used only for visual disambiguation in Figure 1 and for the descriptive analysis of intra-guild blame in the Results; it does not affect the calculation of network metrics. One additional node ("The Inquiry's questions") was excluded from the analytical network as it represents a procedural element rather than a responsibility-bearing actor; it appears in Figure 1 for visual completeness only.

### *Network representation and measures*

The network is analysed as a directed, unweighted graph. Directionality reflects the orientation of blame attribution from the attributing actor to the actor being blamed. The network does not include signed or weighted edges, as the Inquiry synthesis does not provide a consistent basis for distinguishing the magnitude or polarity of blame beyond its presence or absence.

To address the first objective, a set of standard network measures commonly used in the analysis of directed social networks was computed, following established formulations in social network analysis and network science (Wasserman & Faust, 1994; Freeman, 1979; Holland & Leinhardt, 1976; Newman, 2002, 2006, 2010; Scott, 2017; Brandes et al., 2013).

At the node level, out-degree and in-degree were calculated to characterise how blame is distributed across actors and to identify those that emerge as focal points of blame attribution. These measures capture the extent to which actors direct blame toward others and the extent to which they become the targets of responsibility claims.

At the network level, degree centralisation was calculated to assess the overall concentration of blame attribution across the network, while reciprocity was used to examine the symmetry or asymmetry of blame relationships. Together, these measures provide a structural characterisation of the distribution and directionality of blame, directly addressing the first objective of the study.

To further characterise the organisation of blame attribution, transitivity (or clustering) was computed to describe the tendency for blame relationships to form closed triads (Watts & Strogatz, 1998; Newman, 2003; Fagiolo, 2007). This measure captures patterns of local organisation in blame attribution without implying coordination, coalition formation, or shared intent among actors. Community structure was assessed using modularity and the Walktrap algorithm (Newman, 2006; Pons & Latapy, 2005).

### *Reference models*

To address the second objective of the study, the observed blame attribution network was compared with a set of reference models designed to establish neutral benchmarks for structural comparison (Butts, 2009; Hobson et al., 2021). The models form a sequence of null hypotheses, each controlling for different levels of structural constraint. This allows us to progressively test whether observed network features require explanation beyond increasingly complex baseline assumptions.

Reference or null models are widely used in network analysis to provide structural baselines against which observed patterns can be evaluated, without implying specific behavioural mechanisms or strategic decision rules (Butts, 2009; Snijders, 2011; Newman, 2010). In this study, reference models are used strictly as comparative benchmarks to assess whether observed network properties depart from expectations under simple structural constraints. Deviations from these baselines are therefore interpreted as evidence of patterned organisation in the distribution of blame, rather than as direct indicators of actor intent, coordination, or strategic behaviour. Accordingly, we implement multiple reference models, each preserving different structural properties of the observed network.

1. Random directed networks (Erdős–Rényi). Ensembles of random directed networks with the same number of nodes and edges as the observed network were generated (Erdős & Rényi, 1960). This model provides a minimal baseline, testing whether the observed structure differs from what would be expected if every possible accusation were equally likely, with no actor-level tendencies.

2. Out-degree–preserving random networks. Ensembles were generated using a randomisation algorithm that retains each actor’s empirical out-degree (i.e., the number of others each actor blames) while randomly reassigning the targets of those accusations. This model controls for heterogeneity in how prolifically different actors attribute blame, testing whether structure can be explained solely by differences in actors’ propensity to accuse others.

3. Degree-preserving random networks (directed configuration model). Ensembles were generated using a directed configuration model that retains the empirical sequences of both in-degree and out-degree while randomising edge placement (Newman, 2010). This is the most stringent baseline used, as it preserves the exact distributions of both blaming and being blamed. It tests whether higher-order structural features (such as reciprocity or clustering) remain notable once the fundamental activity and popularity of each actor are accounted for.

4. Preferential-attachment networks. As a commonly used generative benchmark in social network analysis, ensembles were generated via a directed preferential-attachment process (Barabási & Albert, 1999). This model produces networks where edges are more likely to be directed toward already popular nodes (high in-degree). It is included not as a hypothesized mechanism for blame attribution, but as a well-known social-structural benchmark for comparison.

For each reference model, network measures were computed across a large number of simulated networks (999 replicates per model), generating reference distributions against which the observed values were compared. These comparisons provide the empirical basis for assessing whether the observed structural features of the blame attribution network reflect non-random organisation that cannot be reduced to simple compositional constraints, thereby addressing the second objective of the study.

### *Analytical workflow and software implementation*

All analyses were conducted using established network analysis software (Butts, 2008; Csárdi & Nepusz, 2006). Network construction, calculation of descriptive measures, and generation of reference models were implemented as described below.

All analyses were performed in R (R Core Team, 2023). Network construction, descriptive statistics, and visualization were carried out using the igraph package (Csárdi & Nepusz, 2006). Reference networks were generated using a combination of igraph functions (including rewire and erdos.renyi.game) and custom scripts designed to implement the specified randomisation constraints. Statistical analyses, including Kruskal–Wallis and Wilcoxon tests, were conducted using the stats package included in base R. Outlier detection in the relationship between degree and betweenness centrality was performed using the outlierTest function from the car package (Fox & Weisberg, 2019).

Figures were produced using a combination of igraph plotting functions and the ggplot2 package (Wickham, 2016). Network visualisations employed the Kamada–Kawai layout algorithm to represent structural relationships among actors. All network data, digitisation procedures, and analysis code are provided as Supplementary Material.

## Results

### *Structural properties of the Grenfell blame network*

The analytical network comprised 21 nodes and 83 directed edges (Fig. 1). Network density was 0.198, meaning that about one in five of all possible accusations was present (Table 1). The diameter was 3 and the average path length 1.68, indicating that most actors were separated by only one or two steps. Reciprocity was 0.48, meaning that nearly half of all accusations were reciprocated. The clustering coefficient (transitivity) was 0.621, indicating that blame attributions tended to form tightly knit triads (closed triangles) rather than occurring in isolated dyads. Degree assortativity, measured from out-degree to in-degree, was negative (–0.27), suggesting that actors who directed many accusations tended to target actors that were less frequently blamed. Degree entropy was 2.656, reflecting heterogeneity in how accusations were distributed. Community detection using the Walktrap algorithm assigned all nodes to a single group, yielding modularity = 0.000 (Table 1).

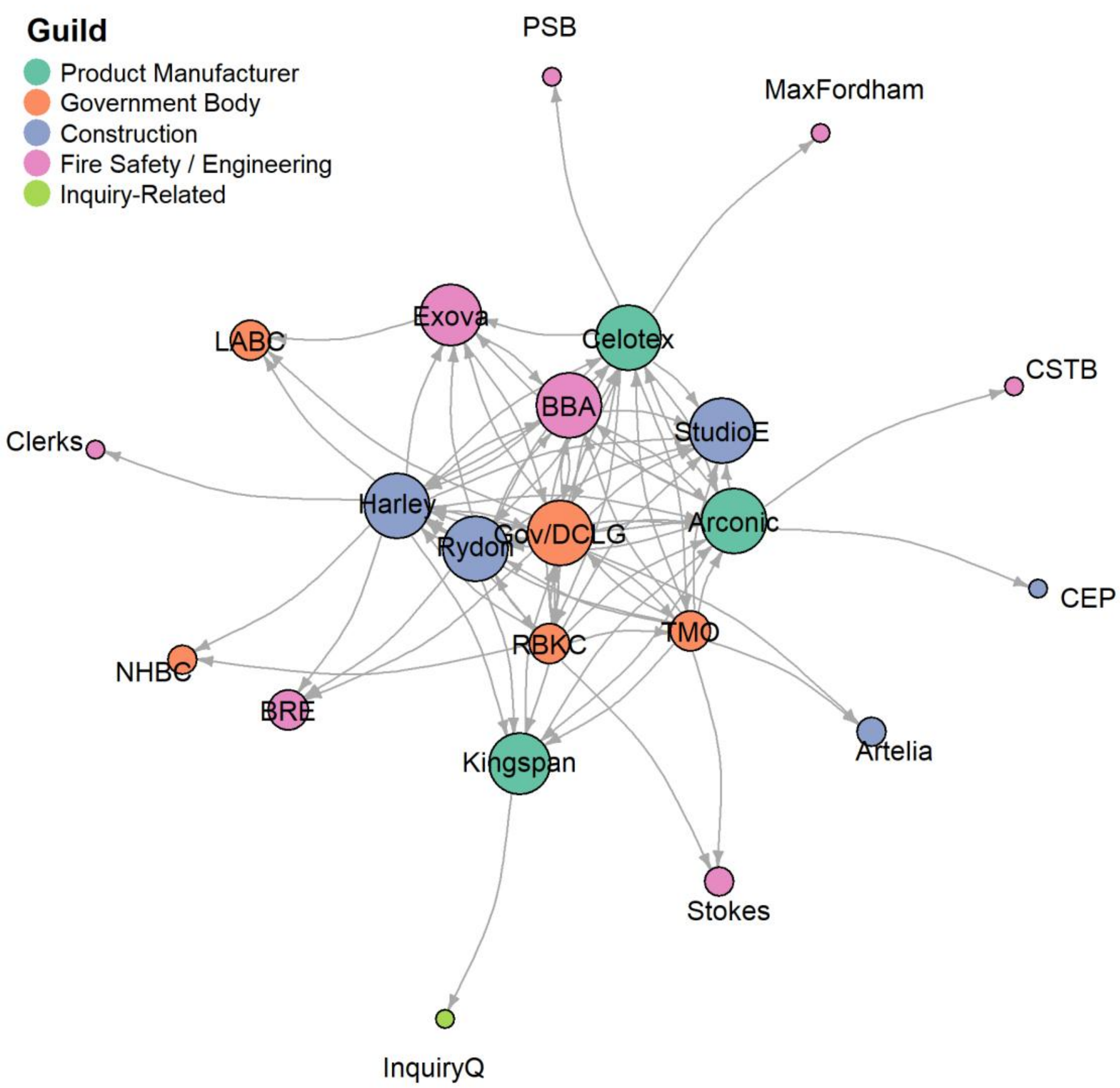

**Figure 1.** Visualization of the Grenfell blame network. Node size scales with in-degree (the number of incoming blame attributions). Node colour corresponds to the actor's guild membership (see Methods for definitions), and arrows represent directed blame attributions.

**Table 1.** Structural metrics of the Grenfell blame network.

| Metric | Value |
|---|---|
| Number of Nodes | 21 |
| Number of Edges | 83 |
| Diameter | 3 |
| Average Path Length | 1.68 |
| Reciprocity | 0.48 |
| Assortativity | -0.27 |
| Modularity | 0 |
| Transitivity | 0.621 |

### *Comparison with reference models*

The four reference models differed markedly in their ability to reproduce the observed structural properties of the Grenfell blame network (Fig. 2). For assortativity, the degree-preserving model reproduced the strongly negative value observed in the empirical network. Out-degree–preserving randomizations shifted the coefficient upward toward zero, but their simulated distribution still encompassed the observed value. In contrast, preferential-attachment and random models significantly underestimated the magnitude of assortativity. For average path length, both the degree-preserving and out-degree–preserving models closely matched the observed value, whereas random graphs and preferential-attachment models overestimated it. With respect to clustering, only the degree-preserving model captured the high clustering coefficient, which was underestimated by the other models. Degree entropy was exactly reproduced by the degree-preserving model, as this property is fixed by construction, while the other three models underestimated it. Modularity was only partially reproduced by the degree-preserving model: its simulated distribution had a mean slightly above zero but a confidence interval that included the observed value of 0. By contrast, the out-degree–preserving, preferential-attachment, and random models significantly overestimated modularity. Finally, reciprocity was accurately reproduced by the degree-preserving model, while the other three models significantly underestimated it.

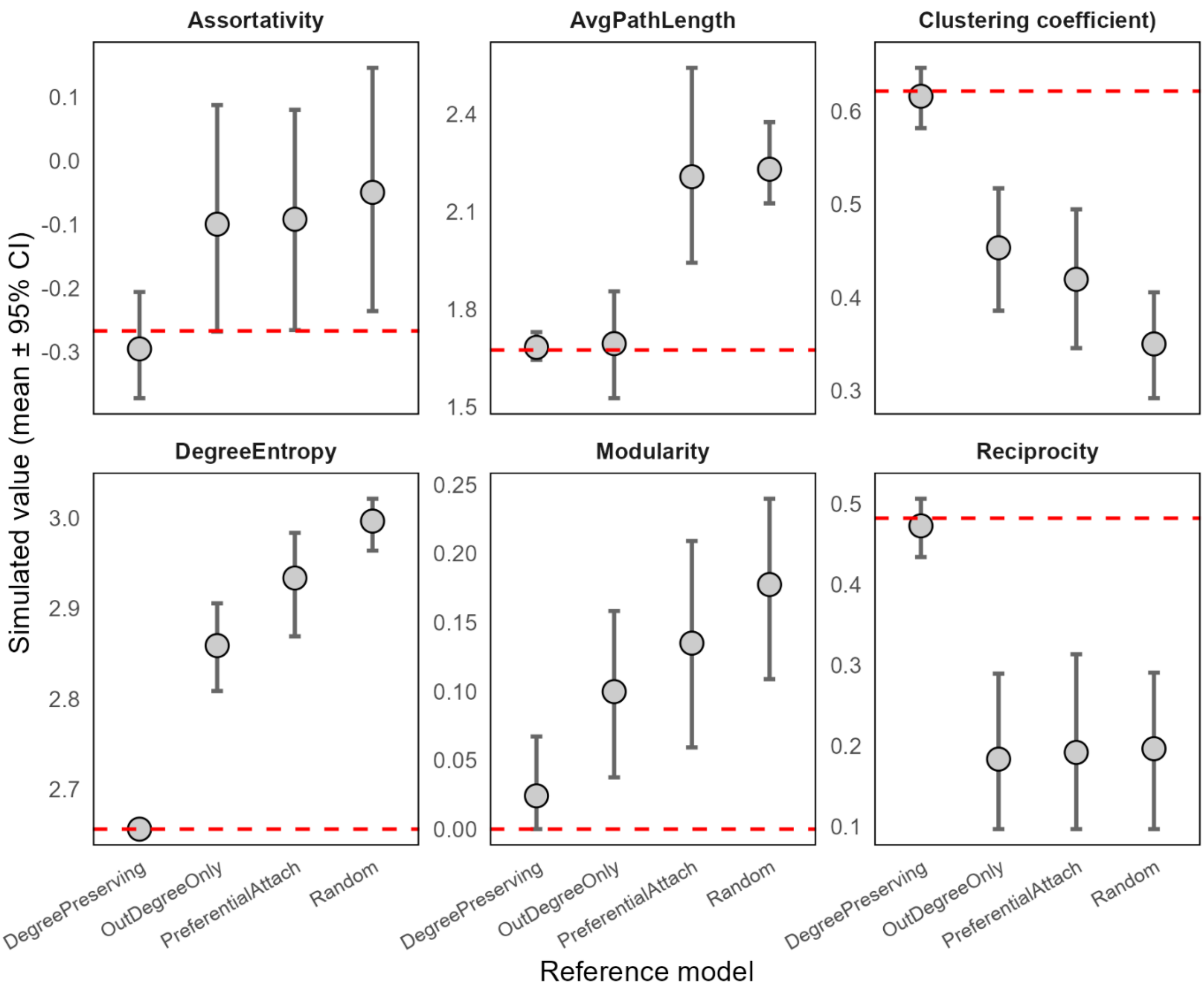

**Figure 2.** Comparison of observed network metrics with expectations from four alternative reference models. Points indicate the mean value of each reference model ensemble (999 replicates), with vertical bars showing the 95% interval of simulated values. The horizontal dashed red line marks the value observed in the Grenfell blame network.

The multivariate analysis showed the degree-preserving model yielded the smallest distances to the observed network (mean = 0.067, SD = 0.026). Out-degree–preserving randomizations showed significantly larger deviations (0.462 ± 0.063), followed by preferential-attachment simulations (0.745 ± 0.126), while purely random graphs produced the largest departures (0.820 ± 0.057). The Kruskal–Wallis test confirmed strong overall differences among models ($p < 0.001$), and pairwise Wilcoxon tests with Holm correction indicated that all four ensembles differed significantly from each other (all adjusted $p < 0.0001$; Fig. 3).

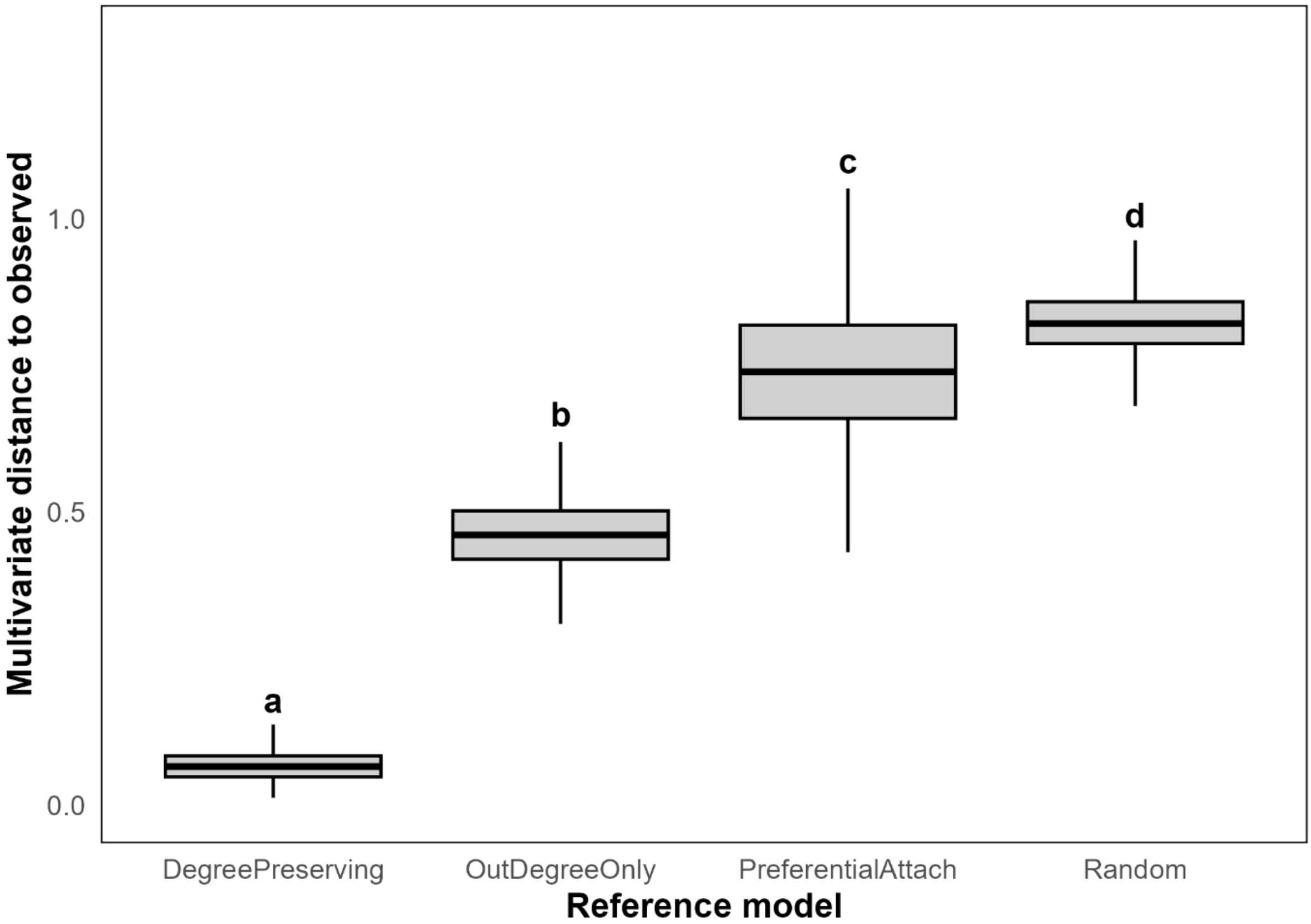

**Figure 3**. Multivariate Euclidean distance between the observed Grenfell network and four reference models, calculated across multiple network metrics. Boxplots show the distribution of distances across 999 randomizations per model. Models with different letters differ significantly according to pairwise Wilcoxon rank-sum tests with Holm correction ($p < 0.05$).

Regarding the analysis of whether intraguild accusations occurred more frequently than expected by chance, 19% of accusations in the observed network were directed within the same guild. This proportion was close to the null expectation of 22% obtained by random shuffling of guild labels (999 replicates) and fell well within the expected range (empirical two-tailed p = 0.37; Fig. 4).

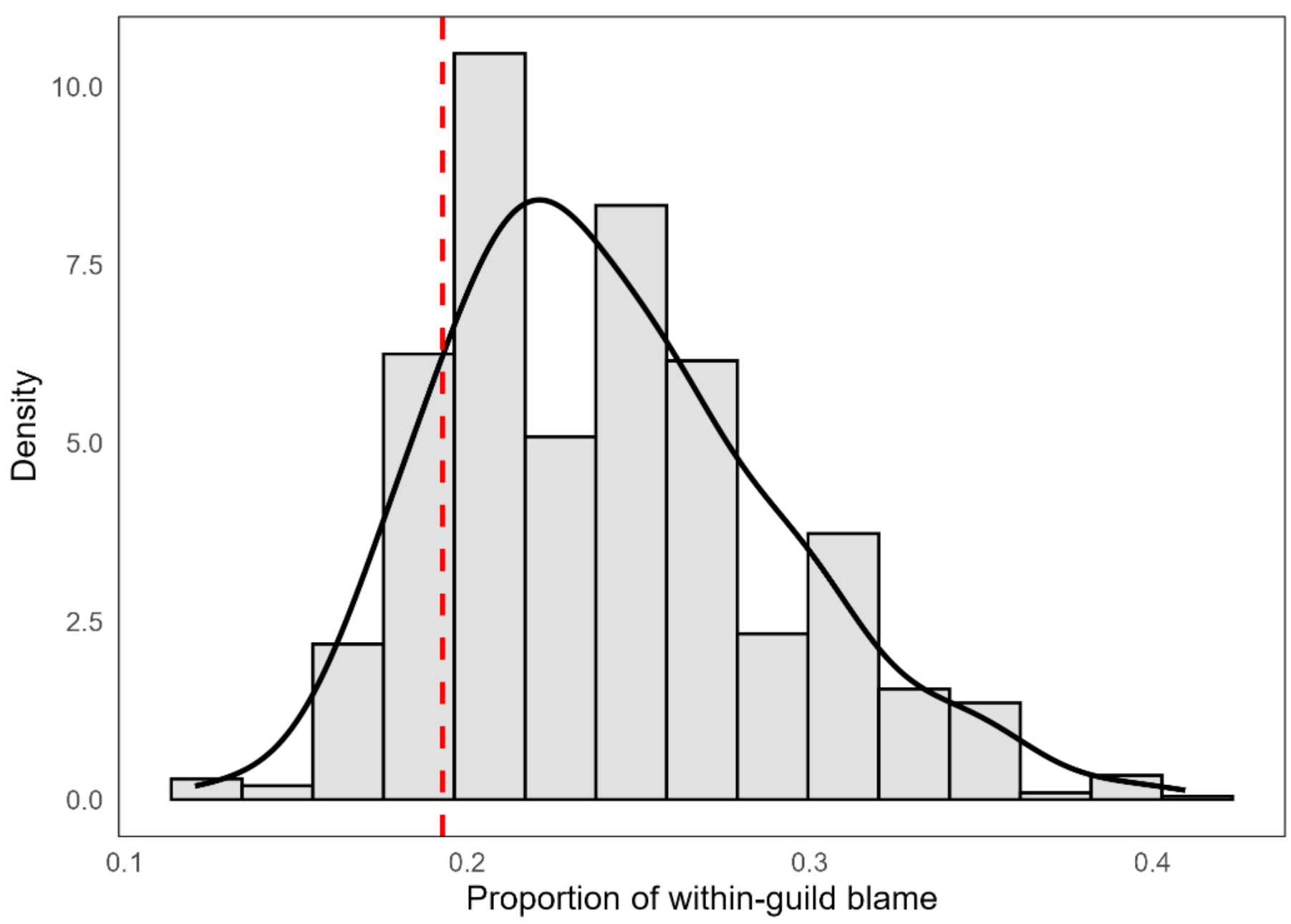

**Figure 4.** Null distribution of the proportion of intra-guild blame, obtained by randomly shuffling guild labels across actors (999 iterations). The histogram (gray bars) shows the simulated distribution, with the black curve indicating its kernel density estimate. The vertical red dashed line marks the observed proportion of intra-guild blame in the Grenfell network.

### *Node-level measures*

Accusations were unevenly distributed across actors. Government/DCLG, Harley Facades, RBKC, and TMO recorded the highest out-degrees, while Studio E, Arconic, and Celotex had some of the highest in-degree values (Fig. 1; Appendix Table S2).

The relationship between in-degree and out-degree in the Grenfell blame network was strongly positive (Spearman's $\rho = 0.68$, $p < 0.001$; Fig. 5). Most actors displayed either simultaneously low in-degree and low out-degree or simultaneously high in-degree and high out-degree. Smaller groups of actors fell outside this diagonal pattern, showing either high out-degree with low in-degree or low out-degree with high in-degree. Very few actors were located near the average values of in-degree or out-degree.

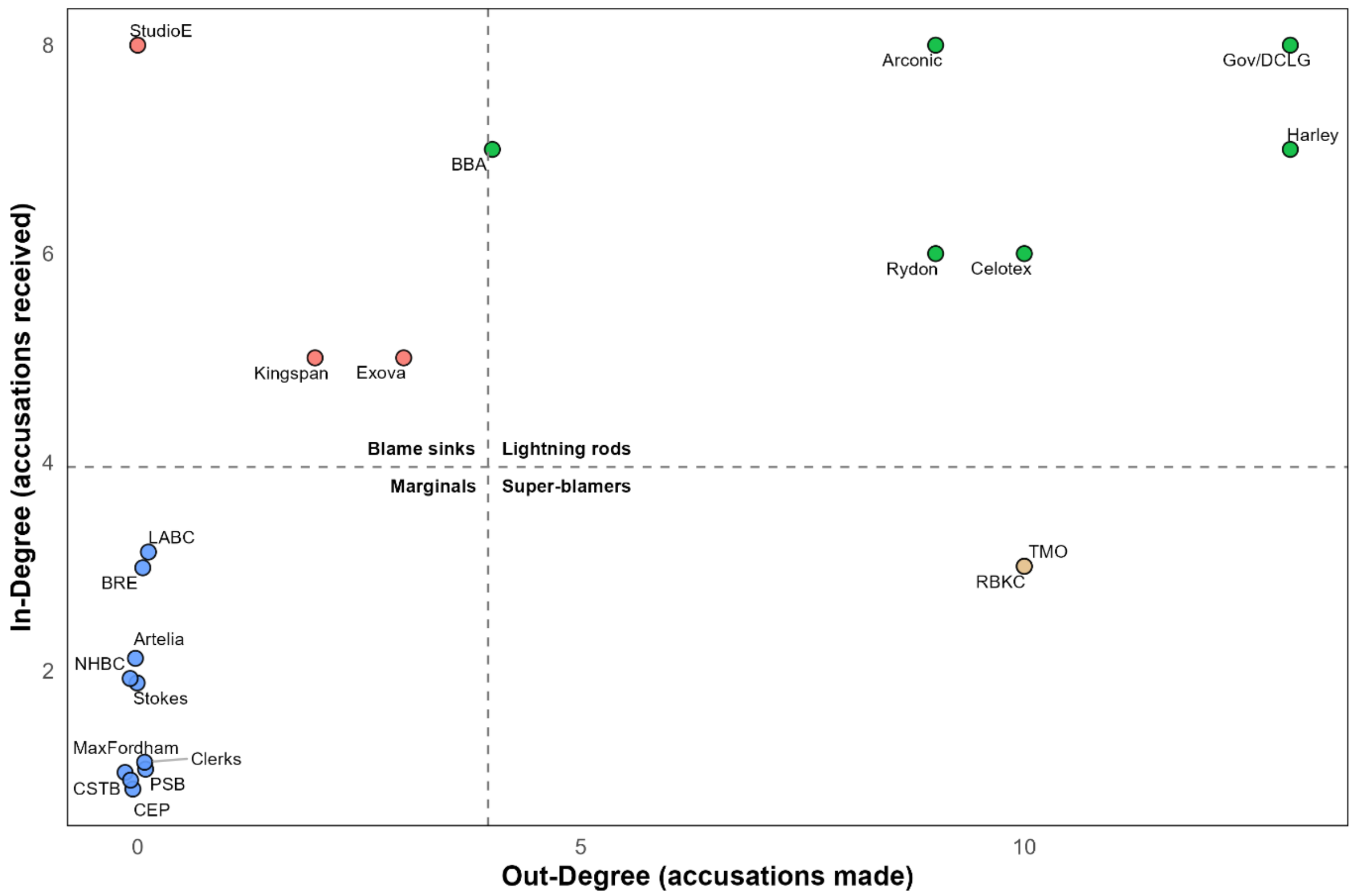

**Figure 5.** Relationship between out-degree and in-degree of nodes in the Grenfell blame network. Each point represents one or more nodes, with the size of the point indicating the number of overlapping cases (i.e., nodes sharing the same in-degree and out-degree values). Labels identify overlapping nodes by their numeric IDs.

The relationship between total degree and normalized betweenness centrality in the Grenfell blame network was positive (Spearman's $\rho = 0.70$; $p < 0.001$; Fig. 6), indicating that actors with higher degrees tended to also exhibit higher betweenness. A linear regression further identified one statistically significant outlier, Government/DCLG, whose betweenness was higher than expected for its degree (Bonferroni-adjusted $p = 0.045$). Manufacturers such as Celotex, Arconic, and Harley Facades combined relatively high degrees with moderately elevated betweenness, whereas actors such as LABC, Exova, and Studio E had some degree connections but near-zero betweenness, placing them at the periphery of the network.

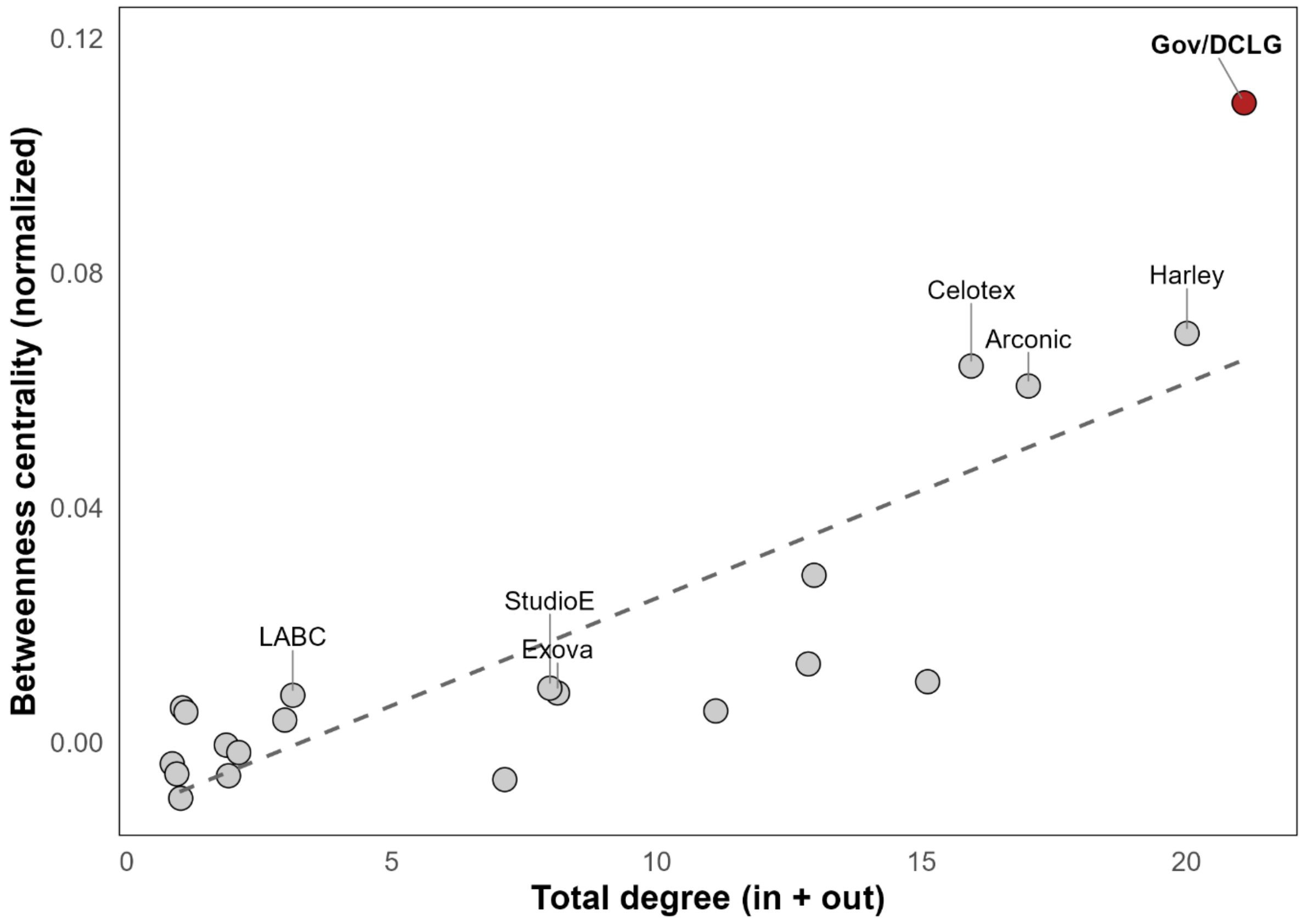

**Figure 6.** Relationship between total degree (in + out) and normalized betweenness centrality for nodes in the Grenfell blame network. The dashed line represents the fitted linear regression. The single red point denotes the node identified as a statistical outlier ($p < 0.05$, Bonferroni corrected) based on studentized residuals. Selected additional nodes (Harley Facades, Arconic, Celotex, LABC, Exova, Studio E) are labelled for reference.

## Discussion

This study contributes to the literature on crisis governance and blame attribution by applying network-analytic tools to the web of blame presented during the Phase 2 closing proceedings of the Grenfell Tower Inquiry. As part of those proceedings, Counsel to the Inquiry assembled a structured representation of how core participants attributed responsibility to one another, offering a rare and explicit institutional articulation of blame relations within a major public inquiry. Building on this conceptualisation, the present analysis examines the structural properties of that web using methods from network science, treating it as an empirical object in its own right.

While research on blame games and accountability has predominantly relied on qualitative case studies, document analysis, or actor-centred interpretations of responsibility shifting (Hood, 2011; Weaver, 1986), a network perspective allows the Inquiry's representation of blame attribution to be examined at the level of the entire institutional field. By moving beyond dyadic accounts of blame assignment, this approach makes it possible to identify structural regularities, such as concentration, reciprocity, and clustering, that are not readily apparent through narrative or actor-by-actor analysis. In this sense, the contribution of the study lies not in proposing new accounts of blame behaviour, but in extending existing work on crisis governance through a systematic relational analysis of institutionalised responsibility claims (Ansell & Boin, 2019).

At the network level, the overall density and short average path lengths indicate that blame attributions were highly interconnected rather than isolated. Responsibility claims circulated within a relatively compact institutional space, in which actors were closely linked through overlapping and mutually reinforcing patterns of blame attribution. Such structural proximity suggests that blame attribution in the Inquiry unfolded within a tightly coupled institutional environment, where responsibility claims were unlikely to remain confined to bilateral exchanges. Comparable patterns of institutional entanglement have been observed in analyses of other large-scale crises, where accountability processes expose dense interdependencies across organisational boundaries (Ansell et al., 2016; Boin & Lodge, 2021).

Reciprocity emerged as a prominent feature of the blame attribution network. A substantial proportion of responsibility claims were mutual, indicating that actors frequently attributed blame to those who also attributed responsibility to them. From a structural perspective, this pattern reflects reciprocal responsibility shifting rather than unilateral assignment of fault. Such reciprocity is consistent with theoretical accounts of blame games, in which actors respond to external accusations by redirecting responsibility toward their accusers, thereby mitigating reputational damage and diffusing accountability pressures (Hood, 2011; Weaver, 1986). Importantly, the presence of reciprocal blame attributions does not imply coordination or strategic intent; rather, it reflects a common structural outcome in institutional settings where multiple actors face simultaneous exposure to scrutiny.

The analysis also revealed substantial local clustering, with blame attributions forming tightly connected triads and subgroups. This clustering indicates that responsibility claims tended to concentrate within subsets of actors linked by regulatory, contractual, or governance relationships. Such local organisation is consistent with the institutional architecture of the Grenfell case, in which housing provision, regulation, certification, and oversight were distributed across interconnected public and private entities. From a governance perspective, these clustered patterns illustrate how accountability processes often mirror existing institutional arrangements, producing responsibility structures that align with organisational boundaries and oversight chains rather than cutting cleanly across them (Ansell et al., 2016; Boin & 't Hart, 2010).

At the same time, the observed modularity of the network indicates that blame attribution was not evenly distributed across the institutional landscape. Distinct clusters of actors were more densely connected internally than externally, suggesting that responsibility claims were often articulated within relatively bounded institutional domains. This finding aligns with research on institutional crises, which highlights how accountability processes may fragment along sectoral or organisational lines even when failures are systemic in nature (Ansell et al., 2016). The coexistence of dense clustering and modular structure underscores the tension between systemic failure and segmented accountability that characterises many complex crises.

Actor-level analyses further illustrate how these structural dynamics played out unevenly across participants. A small number of organisations occupied highly central positions in the blame attribution network, attracting a disproportionate share of incoming responsibility claims. These actors emerged as focal points of blame attribution, not necessarily because they exercised greater control or intent, but because of their formal roles within the governance system and their visibility within the Inquiry process. High betweenness centrality for certain actors reflects structurally central positions in the circulation of blame attribution, indicating their location at the intersection of multiple responsibility pathways.

The comparison with reference models reinforces the interpretation of these findings as structurally distinctive but not behaviourally determinative. Observed levels of reciprocity, clustering, and centralisation departed from those expected under random and degree-preserving benchmark models (Wasserman & Faust, 1994; Newman, 2010), indicating that simple structural constraints alone cannot account for the configuration of the blame attribution network. However, deviations from reference models do not demonstrate intentional coordination, strategic behaviour, or dynamic interaction. Rather, they indicate that the observed structure reflects institutional regularities and patterned accountability relationships embedded in the governance context of the Inquiry, consistent with accounts of blame avoidance and institutional crisis in complex governance settings (Hood, 2011; Ansell et al., 2016).

These findings contribute to the literature on blame avoidance and crisis governance by demonstrating how network analysis can be used to describe the relational structure of accountability processes without exceeding the epistemic limits imposed by static data. Whereas much existing work on blame games relies on qualitative reconstructions of strategic interaction (Hood, 2011; Weaver, 1986), the present study shows that network methods can complement such approaches by identifying structural regularities in how responsibility is publicly articulated within formal institutional arenas. In this sense, the analysis does not reveal hidden intentions or strategies, but provides a systematic description of the relational environment in which blame attribution unfolds.

More broadly, the Grenfell case illustrates how institutional arenas of inquiry can generate patterned distributions of responsibility that shape post-crisis accountability debates. By formalising and aggregating blame attributions, the Inquiry produced a structured representation of responsibility that both reflects and reinforces existing institutional arrangements. Such representations may influence subsequent political, legal, and organisational responses by foregrounding certain actors and relationships while marginalising others. Understanding the structural properties of these representations is therefore essential for analysing how accountability is constructed and contested in the aftermath of major crises.

Several important limitations delineate the scope of our analysis and shape the interpretation of our findings. First, the network constitutes a static snapshot of blame attributions synthesised by the Inquiry at a single procedural moment; it does not capture the dynamic, iterative processes through which responsibility claims may have evolved over the course of the Inquiry or within the wider public sphere. This static perspective, while appropriate for structural description, cannot reveal sequences, cascades, or adaptive responses over time (Snijders, 2011). Second, the boundaries of the network are defined by the Inquiry's own procedural and illustrative synthesis. Alternative boundary specifications, for example those incorporating victim advocacy groups, specialist subcontractors, media actors, or political institutions, would necessarily yield different network configurations and could alter observed properties such as centrality or modularity (Borgatti & Halgin, 2011). The analysis therefore describes the structure of blame attribution as formalised by the Inquiry, rather than an exhaustive or objective map of all possible responsibility relations. Third, while the analysis identifies and quantifies salient structural patterns, including high levels of reciprocity and clustering, it does not permit attribution of these patterns to specific underlying mechanisms, such as deliberate strategic behaviour, coalition formation, or private intent. The data and methods are descriptive; examining causal processes would require complementary longitudinal, ethnographic, or interview-based approaches (Hobson et al., 2021).

Taken together, these limitations define the nature and value of the contribution offered here. By providing a systematic and reproducible structural description of the blame attribution network as formalised by a major public inquiry, the study establishes a quantitative baseline against which theoretical accounts of blame games can be assessed and offers a methodological template for analysing accountability discourses in other institutional crises.

Finally, while this study focuses on a single case, the approach has wider relevance for the study of governance under conditions of uncertainty and institutional complexity. Network-based analyses of blame attribution could be applied to other public inquiries, regulatory failures, or crisis investigations to compare how different institutional settings shape patterns of responsibility. Such comparative work would not aim to infer strategy or causality from network structure alone, but to examine how formal accountability processes produce relational configurations of blame that condition subsequent governance outcomes.

## References

Ansell, C., & Boin, A. (2019). Taming deep uncertainty: The potential of pragmatist principles for understanding and improving strategic crisis management. *Administration & Society, 51*(7), 1079–1112. https://doi.org/10.1177/0095399717747655

Ansell, C., Boin, A., & Kuipers, S. (2016). Institutional crises and the mobilization of governance capacity. *Governance, 29*(2), 195–217. https://doi.org/10.1111/gove.12164

Barabási, A.-L., & Albert, R. (1999). Emergence of scaling in random networks. *Science, 286*, 509–512.

Beck, U. (1992). *Risk society: Towards a new modernity*. SAGE Publications.

Boin, A., & Lodge, M. (2021). Designing resilient institutions for transboundary crisis management: A time for public administration. *Public Administration, 99*(3), 429–442. https://doi.org/10.1111/padm.12719

Boin, A., & 't Hart, P. (2010). Organising for effective emergency management: Lessons from research. Australian Journal of Public Administration, 69(4), 357–371. https://doi.org/10.1111/j.1467-8500.2010.00694.x

Borgatti, S. P., & Halgin, D. S. (2011). On network theory. *Organization Science*, 22, 1168–1181. https://doi.org/10.1287/orsc.1100.0641

Brandes, U., Robins, G., McCranie, A., & Wasserman, S. (2013). What is network science? *Network Science, 1*(1), 1–15. https://doi.org/10.1017/nws.2013.2

Butts, C. T. (2009). Revisiting the foundations of network analysis. *Science, 325*(5939), 414–416. https://doi.org/10.1126/science.1171022

Csárdi, G., & Nepusz, T. (2006). The igraph software package for complex network research. *InterJournal, Complex Systems*, 1695.

Erdős, P., & Rényi, A. (1960). On the evolution of random graphs. *Publications of the Mathematical Institute of the Hungarian Academy of Sciences, 5*, 17–61.

Fagiolo, G. (2007). Clustering in complex directed networks. *Physical Review E, 76*(2), 026107. https://doi.org/10.1103/PhysRevE.76.026107

Freeman, L. C. (1979). Centrality in social networks: Conceptual clarification. *Social Networks, 1*(3), 215–239. https://doi.org/10.1016/0378-8733(78)90021-7

Grenfell Tower Inquiry. (2024). *Counsel to the Inquiry's closing submissions: Web of blame diagram*. Grenfell Tower Inquiry. https://www.grenfelltowerinquiry.org.uk

Heinkelmann-Wild, T., & Zangl, B. (2020). Multilevel blame games: Blame-shifting in the European Union. *Governance, 33*(4), 953–969. https://doi.org/10.1111/gove.12446

Hinterleitner, M., & Sager, F. (2015). Avoiding blame – A comprehensive framework and the Australian Home Insulation Program fiasco. *Policy Studies Journal, 43*(1), 139–161. https://doi.org/10.1111/psj.12086

Hobson, E. A., Silk, M. J., Fefferman, N. H., Larremore, D. B., Rombach, P., Shai, S., & Pinter-Wollman, N. (2021). A guide to choosing and implementing reference models for social network analysis. *Biological Reviews, 96*(6), 2716–2734. https://doi.org/10.1111/brv.12775

Holland, P. W., & Leinhardt, S. (1976). Local structure in social networks. *Sociological Methodology, 7*, 1–45. https://doi.org/10.2307/270703

Hood, C. (2011). *The blame game: Spin, bureaucracy, and self-preservation in government*. Princeton University Press.

Hood, C. (2014). The politics of blame avoidance in public policy. *British Journal of Political Science, 44*(4), 635–655. https://doi.org/10.1017/S0007123411000295

Lukes, S. (2005). *Power: A radical view* (2nd ed.). Palgrave Macmillan.

Newman, M. E. J. (2002). Assortative mixing in networks. *Physical Review Letters, 89*(20), 208701. https://doi.org/10.1103/PhysRevLett.89.208701

Newman, M. E. J. (2006). Modularity and community structure in networks. *Proceedings of the National Academy of Sciences, 103*(23), 8577–8582. https://doi.org/10.1073/pnas.0601602100

Newman, M. E. J. (2010). *Networks: An introduction*. Oxford University Press.

Pons, P., & Latapy, M. (2005). Computing communities in large networks using random walks. In *Computer and Information Sciences – ISCIS 2005* (pp. 284–293). Springer. https://doi.org/10.1007/11569596_31

R Core Team. (2023). *R: A language and environment for statistical computing*. R Foundation for Statistical Computing.

Roulet, T. J., & Pichler, R. (2020). Blame game theory: Scapegoating, whistleblowing and discursive struggles following accusations of organizational misconduct.

*Organization Theory, 1*(4), 2631787720982514. https://doi.org/10.1177/2631787720982514

Scott, J. (2017). *Social network analysis* (4th ed.). SAGE Publications.

Snijders, T. A. B. (2011). Statistical models for social networks. *Annual Review of Sociology, 37*, 131–153. https://doi.org/10.1146/annurev.soc.012809.102709

Ulrich, W., & Gotelli, N. J. (2007). Null model analysis of species nestedness patterns. *Ecology, 88*(7), 1824–1831. https://doi.org/10.1890/06-1205.1

Wasserman, S., & Faust, K. (1994). *Social network analysis: Methods and applications*. Cambridge University Press. https://doi.org/10.1017/CBO9780511815478

Watts, D. J., & Strogatz, S. H. (1998). Collective dynamics of “small-world” networks. *Nature, 393*(6684), 440–442. https://doi.org/10.1038/30918

Weaver, R. K. (1986). The politics of blame avoidance. *Journal of Public Policy, 6*(4), 371–398. https://doi.org/10.1017/S0143814X00004219

Wickham, H. (2016). *ggplot2: Elegant graphics for data analysis*. Springer.